\documentclass[12pt]{article}

\newcommand{\beq}{\begin{equation}}
\newcommand{\eeq}{\end{equation}}
\newcommand{\be}{\begin{equation}}
\newcommand{\ee}{\end{equation}}
\newcommand{\beqn}{\begin{eqnarray}}
\newcommand{\eeqn}{\end{eqnarray}}
\newcommand{\bea}{\begin{eqnarray}}
\newcommand{\eea}{\end{eqnarray}}

\DeclareFixedFont{\xiiss}{OT1}{cmss}{m}{n}{12}
\DeclareFixedFont{\ixss}{OT1}{cmss}{m}{n}{9}
\DeclareFixedFont{\cmrnine}{OT1}{cmr}{m}{n}{9}

\newcommand{\CC}{\hbox{\xiiss C\kern-.4emI}}
\newcommand{\RR}{\hbox{\xiiss R\kern-.45emI}}
\newcommand{\ZZ}{\hbox{\xiiss Z\kern-.4emZ}}
\newcommand{\CCs}{\hbox{\ixss C\kern-.4emI}}
\newcommand{\ZZs}{\hbox{\ixss Z\kern-.4emZ}}
\newcommand{\pa}{\partial}

\newcommand{\pasl}{\pa\kern-.55em /}
\newcommand{\Dsl}{D\kern-.65em /}

\def\href#1#2{#2}

\begin{document}
\begin{titlepage}
\title{ 
        \begin{flushright}
        \begin{small}
        RU-NHETC-99-38\\
        hep-th/9910238\\
        \end{small}
        \end{flushright}
        \vspace{1.cm}
Hypermultiplet Moduli Space and Three Dimensional Gauge Theories
 }

\author{
Moshe Rozali\thanks{e-mail: \tt rozali@physics.rutgers.edu}\\
\\
        \small\it Department of Physics and Astronomy\\
        \small\it Rutgers University\\
        \small\it Piscataway, NJ 08855
}

\maketitle

\begin{abstract}

We establish a relation, conjectured recently by E. Witten, between the hypermultiplet moduli space in compactifications of the heterotic string
on an A-D-E singularities, and the moduli spaces of  three dimensional  pure gauge theories with the corresponding A-D-E gauge groups. It is 
possible to add a bounded number of heterotic fivebranes sitting in the singularity, while (in leading order in $\alpha'$) keeping the heterotic string perturbative. The corresponding hypermultiplet moduli space is given by
the moduli space of a three dimensional gauge theory with matter.

\end{abstract}
\end{titlepage}


\section{Introduction}

The duality between the heterotic string on $K3 \times T^2$ and type II string theory on CY manifolds \cite{dual} can be used to study the moduli space of four  dimensional
string vacua with 8 supercharges. The moduli space factorizes  locally into 
the hyper- and vector multiplet moduli spaces. It is then possible to study 
the hypermultiplet moduli space classically in the heterotic string side, since the heterotic string dilaton is a member of a vector multiplet. This is to be compared with calculations on the type II side, as in \cite{typeII}.

The detailed study of singularities in the moduli space is, however, not always possible
at tree level.
 There  are some singularities that drive the string coupling to diverge, no matter how small it is asymptotically. In those cases the metric on moduli space can be computed reliably, but the physics of the singularity is non-perturbative. Well-known singularities of that sort are the type II conifold  \cite{conifold}, and the non-perturbative gauge symmetry at the core of a small instanton \cite{instanton}.

In \cite{witten}, a detailed study of singularities which do not  
involve non perturbative string  physics was initiated. An example is the heterotic string near an A-D-E type singularity of a $K3$ surface. Since the heterotic string dilaton becomes weak near the singularity, one may study the singularity in the heterotic CFT.  Indeed, for the case of $A_1$ singularity, the moduli space was found to be the Atiyah-Hitchin manifold. The classical singularity is resolved by a combination of one loop and worldsheet instanton effects, with no recourse to stringy corrections.

The pattern in which the $\alpha'$ corrections smooth out the classical 
singularity is reminiscent of the work in \cite{sw}.  This analogy 
suggested a relation between the two frameworks \cite{witten}\footnote{
The relation between the hypermultiplet moduli space and three dimensional
gauge theory was noted in the type II context in \cite{ss}.}. Therefore it was
conjectured in \cite{witten} that the hypermultiplet
moduli space of the heterotic compactification on an A-D-E type singularity is identical to the moduli space of a three dimensional pure gauge thery with 8 
supercharges and the  corresponding A-D-E  gauge group.

In this note we establish this relation by using the duality between the heterotic string on $T^3$ and M-theory on $K3$. Furthermore, one can construct  slightly more general backgrounds which keep the heterotic coupling perturbative (in leading order in $\alpha'$). This is done by putting a number of heterotic fivebranes at the A-D-E singularity. This does not break the supersymmetry further. In order for the heterotic string coupling to be small near the singularity,  the number of fivebranes is bounded, as discussed below.  The hypermultiplet
moduli space in this case turns out to be the moduli space of a three dimensional gauge theory with matter.

 Dualities in three dimensional gauge theories  exchanging the Coulomb and Higgs branches were discussed in \cite{is}, and were 
named Mirror symmetry. Realization of the mirror symmetry
by an embedding in string theory \cite{mirror} has concentrated mainly on 
realizing the gauge theory on D-branes (see, however, \cite{pz}  for a closely related discussion). 
 Here, the embedding in M-theory (or string theory) involves the dynamics of coincident membranes in M-theory, or coincident 5-branes (compactified on $T^3$) in the heterotic string theory.

 As is shown below in a particular case, the requirement of the heterotic string to be perturbative is a necessary condition for the absence of singularities in the hypermultiplet moduli space. It  would be interesting to verify that this is also a sufficient condition, along the lines of \cite{witten}, or using the gauge theory representation. For $A_N$ singularities with no small instantons, there is 
an  independent argument for the smoothness (and identification) of the hypermultiplet
moduli space \cite{sen}.

\section {Heterotic String on A-D-E Singularities}

We consider a compactification of the heterotic string on $K3$, in the limit where the $K3$ developes an A-D-E type singularity, associated with a gauge group which we call $G$. Concentrating on the behavior near the singularity allows one to replace the $K3$ surface by a non-compact ALE space which looks
identical near the singularity. One therefore is allowed to ignore constraints of tadpole cancellation which arise upon further compactification to lower dimensions \cite{svw}. We call this  ALE  space $X_G$.

Concentrating on the singularity results in a  low energy six dimensional 
theory which decouples from gravity.  The hypermultiplet moduli space
is an hyperKahler manifold in this limit. Upon further toroidal 
compactification,
The moduli space factorizes into the vector- and the hypermultiplet moduli spaces. This factorization follows 
from the different R-symmetry transformation laws
for the scalars in the corresponding multiplets.

We are interested in particular in compactifying further on $T^3$, yielding
a three dimensional theory with 8 supercharges. The low energy theory has an R-symmetry
$SO(4) = SU(2)_L \times SU(2)_R$.  The scalars in the vector- and hypermultiplets transform in different $SU(2)$ factors. Therefore the moduli space factorizes.
In particular, the hypermultiplet moduli space is independent of the value of vector moduli, such as the radii of $T^3$ and the heterotic string coupling.
 It is the same in all dimensions \cite{is} and does not receive stringy corrections.

 For  the heterotic  string background discussed here, the three dimensional R-symmetry $SU(2)_L 
\times SU(2)_R$ originates from
the factors $T^3$ and $X_G$, respectively. The scalars in the hypermultiplet moduli space are distinguished by a transformation law under $SU(2)_R$, that is the one that rotates the 3 complex structures of $X_G$.

Since one can safely go to strong heterotic coupling, it is natural to 
consider the situation in dual variables.  In M-theory variables the background in question is $K3 \times X_G$. The first factor comes from the duality 
between M-theory on $K3$ and the heterotic string on $T^3$, and the second factor is the  ALE space introduced above. Since the moduli space in question does not depend on the moduli of $K3$, we can take it to be 
a generic, non-singular surface.

M-theory near an A-D-E singularity is well known to yield a seven dimensional gauge theory with a gauge group $G$ \cite{various}. Further reduction on a smooth
$K3$  surface (with a trivial gauge bundle)
 yields in low energies a pure gauge theory in three dimensions, 8 supercharges, and gauge group G.  

The scalars in the vector multiplet transform  under the R-symmetry $SU(2)$ which originates from $X_G$. 
This identifies the hypermultiplet moduli space on the heterotic side, with the
moduli space of the three dimensional gauge theory with gauge group $G$.

The analogy between the calculations in \cite{witten} and \cite{sw} can be explained by this identification. For example an heterotic worldsheet instanton is an
elemantary string wrapped around a 2-cycle in $X_G$, which we call $\Sigma$. This maps to an $M5$-brane
wrapping $K3$ and the 2-cycle $\Sigma$.  This 5-brane is a magnetic source for the 2+1 dimensional gauge field which comes from the 11 dimensional  3-form reduced on the harmonic
2-form  Poincare dual to $\Sigma$.  In the gauge theory therefore one gets the usual instantonic monopole, which corrects the moduli space metric, as in \cite{sw}. 

\section{Small Instantons and Matter}

In \cite{witten}, the question of resolutions of classical singualrities by 
the heterotic CFT was investigated. One then is interested in classical singularities
which do not drive the heterotic string coupling to be strong. A  general 
set of examples can be obtained following \cite{witten}.

The condition for the heterotic string to remain perturbative near the singularity can be derived from \cite{witten}, equation (1.1):

\beq
\label{bound}
\triangle \phi =  tr(F_{ij}F^{ij})- tr(R_{ij}R^{ij})
\eeq

This relates the variation of the dilaton to the curvature sources of the metric and gauge bundle near the singularity.  

 The dilaton blows up near a small instanton singularity,  and tends to zero near an A-D-E 
singularity. It is therefore possible to add a number of small instantons sitting at the A-D-E singularity, such that the heterotic coupling does not blow up. The number of small instantons is bounded. This bound can be found by 
integrating  equation (\ref{bound}) around the singularity. The condition for the dilaton not to blow up at the singularity is found to be:

\beq
\label{bound1}
s= N_f-N_c \leq 0
\eeq

 where  the instanton number at the singularity is denoted by $N_f$ (a notation to be explained shortly), and the rank of the A-D-E gauge group by $N_c-1$, ($N_c$ equals the Euler characteristic
  of the corresponding ALE space). 

This bound may be modified by $\alpha'$ corrections to (\ref{bound}), and is therefore only a necessary condition for the string coupling to stay small.
 Whether or not the string coupling is small near the singularity has to be decided dynamically in the worldsheet theory, or alternatively in the gauge theory description.

 With instantons included, the local factorization of the moduli space is less effective. For example, even though the Wilson lines on $T^3$ are vector multiplets in three dimensions, they do affect the hypermultiplet moduli
space. As the Wilson lines are tuned to enhance the gauge symmetry to
any gauge group $H$, the hypermultiplet moduli space contains a part which is the moduli space of $H$ instantons. It is only locally, away fron enhanced symmetry points, that the hypermultiplet moduli space is independent of the Wilson lines.

 For the set of examples defined above one may be able to  study the resolution of the classical singularity by the heterotic CFT. Using the duality above, it is straightforward to map the small instanton to the dual variables. One gets M-theory 
on $K3 \times X_G$ with $N_f$ membranes spanning the 2+1 noncompact directions.
The low energy limit is then a gauge theory with matter, in a limit described below.

  An alternative representation of the membranes is as follows.  The 
7 dimensional theory obtained by concentrating near the singularity is a 
gauge theory with a gauge group $G$. One of the terms in the
action is (normalizing the M-theory membrane tension to 1):

\beq
I =  \frac{1}{8 \pi^2} C \wedge tr(F \wedge F)
\eeq
where C is the M-theory 3-form and F is the seven dimensional gauge field. This term comes from the term $C \wedge dC \wedge dC $ in 11 dimensional  supergravity.
Therefore, gauge instantons in seven dimensions carry membrane charge: the membranes spanning the 2+1 noncompact directions can be represented as $G$-instantons on $K3$.

The low energy theory now has additional fields, representing the moduli of the 
instantons. Those fields are hypermultiplets, identified as such by their
R-symmetry representation. The instantons have a finite size on the Higgs branch, and become membranes on the Coulomb branch of the gauge theory.

 A precise identification of the matter content on the Coulomb branch (where the instantons are of zero size)
requires a better understanding of coincident membranes in M-theory.
 The  cases relevant to the six dimensional heterotic hypermultiplet moduli 
space require   tuning the second $K3$ to an enhanced 
symmetry point before taking the limit corresponding to a decompactification of  the $T^3$
on the heterotic side.  Results relevant to the identification
of the matter content in those cases can be found in \cite{pz,mirror}.

We conclude by commenting on the case of $A_{N_c}$ singularities, with a generic $K3$ surface, and membranes seperated on it. We concentrate on this case in order to 
interpret the bound (\ref{bound1}) in the gauge theory.

   The space $X_G$ can be described locally  by a multi-Taub-NUT metric in the 
limit of $N_c$ coincident centers, since both of these space have an $A_{N_c}$ singularity at the origin. The  Taub-NUT metric is a circle fibration, and one can use the circle to reduce to type IIA theory.  In type IIA language this is a D2-D6 system,  where the D6 brane wraps the $K3$ surface. The gauge theory
on the D2-branes is $U(1)^{N_f}$, with an infinite classical gauge coupling, 
related to the infinite asymptotic radius of the circle fiber in the ALE space.
The gauge theory on the 6-branes is $SU(N_c)$, as before. The 2-6 strings generate $N_f$ hypermultiplets in the fundamental representation of 
$SU(N_c)$.

For the $A_{N_c}$ case, $N_c$ and $N_f$ are the number of colors and flavors in the gauge theory $SU(n_c)$.
 One can then identify the number $s = N_f-N_c$ defined above as half the number of fermionic zero modes in the presence of an $SU(N_c)$ monopole. In \cite{sw},  $s$ was required to be positive on order to have (multi-)
monopole corrections to the metric on the Coulomb branch. Such monopoles in the present case correct the metric of both the $SU(N_c)$ and $U(1)^{N_f}$ factors
,
providing therefore a possibility for smoothing out the complete vector
moduli space.

In the absence of those corrections, the moduli space is singular, and the origin of the Coulomb branch represents a non-trivial conformal field theory.
Returning to the heterotic string frame,
we see that the condition for the heterotic string to be perturbative is complementary to the condition in \cite{sw}.  It is then a necessary condition (at least for the $A_{N_c}$ case) for the absence of singularities on the hypermultiplet moduli space. It is not clear whether this condition also guarantees a smooth moduli space.

\section{Acknowledgments}  

We thank  T. Banks, M.
Douglas, K. Intriligator,  A. Rajaraman , E. Witten for useful conversations, and O. Aharony for collaboration in parts of this paper.

\end{document}